\documentclass[onecolumn]{article}  
\usepackage{graphicx}
\usepackage{ amssymb}
\usepackage[ansinew]{inputenc}
\usepackage{authblk}
\RequirePackage{amsmath}
\DeclareMathOperator\erf{erf}

\begin{document}

\title{Experimental evidence of independence of nuclear de-channeling length on the particle charge sign} 

\author[1]{E. Bagli} 
\author[1]{V. Guidi}
\author[1]{A. Mazzolari}
\author[1]{L. Bandiera}
\author[1]{G. Germogli}
\author[1]{A. I. Sytov}
\author[2,3]{D. De Salvador}
\author[4,5]{A. Berra}
\author[4,5]{M. Prest}
\author[6]{E. Vallazza}

\affil[1]{INFN Sezione di Ferrara, Dipartimento di Fisica e Scienze della Terra, Universit{\`a} di Ferrara Via Saragat 1, 44122 Ferrara, Italy}
\affil[2]{Dipartimento di Fisica e Astronomia, Universit{\`a} di Padova, Via Marzolo 8, 35131 Padova, Italy}
\affil[3]{INFN Laboratori Nazionali di Legnaro, Viale dell'Universit{\`a} 2, 35020 Legnaro, Italy}
\affil[4]{Universit{\`a} dell'Insubria, via Valleggio 11, 22100 Como, Italy}
\affil[5]{INFN Sezione di Milano Bicocca, Piazza della Scienza 3, 20126 Milano, Italy}
\affil[6]{INFN Sezione di Trieste, Via Valerio 2, 34127 Trieste, Italy}

\maketitle

\begin{abstract}
Under coherent interactions, particles undergo correlated collisions with the crystal lattice and their motion result in confinement in the fields of atomic planes, i.e. particle channeling. Other than coherently interacting  with the lattice, particles also suffer incoherent interactions with individual nuclei and may leave their bounded motion, i.e., they de-channel. This latter is the main limiting factor for applications of coherent interactions in crystal-assisted particle steering. We experimentally investigated the nature of dechanneling of 120 GeV/c $e^{-}$ and $e^{+}$ in a bent silicon crystal at H4-SPS external line at CERN. We found out that while channeling efficiency differs significantly for  $e^{-}$ ($2\pm2$ $\%$) and  $e^{+}$ ($54\pm2$ $\%$), their nuclear dechanneling length is comparable, $(0.6\pm0.1)$ mm for $e^{-}$ and  $(0.7\pm0.3)$ mm for $e^{+}$. The experimental proof of the equality of the nuclear dechanneling length for positrons and electrons is interpreted in terms of similar dynamics undergone by the channeled particles in the field of nuclei no matter of their charge.
\end{abstract}

In the last decade, a significant boost to the research on particle-crystal interactions was provided by the fabrication of uniformly bent crystals \cite{guidi:113534} with thickness along the beam suitable for experiments at high-energy \cite{Guidi200540,Mazzolari2013130}. Measurements proved the capability of channeling for manipulation of positively \cite{doi.org/10.1140/epjc/s10052-014-2740-7} and negatively \cite{PhysRevLett.112.135503,PhysRevLett.114.074801} charged particle beams from MeV \cite{PhysRevLett.108.014801} up to hundreds of GeV \cite{PhysRevLett.110.175502,PhysRevLett.115.015503}, and for the generation of intense electromagnetic radiation from sub-GeV  \cite{PhysRevLett.112.254801,PhysRevLett.115.025504} to hundreds-GeV electron beams \cite{PhysRevLett.111.255502}. Moreover, channeling effects were exploited for steering \cite{Elishev1979387}, collimation \cite{Scandale201078} and extraction \cite{AfoninU70JETP} of relativistic beams in circular accelerators, as well as splitting and focusing of extracted beams \cite{Denisov1992382}, leading to the installation of two bent crystals in the Large Hadron Collider (LHC) for collimation purposes \cite{LUA9proposal}. The crystals installed in the LHC were successfully tested at 6.5 TeV/c and proved to reduce the beam losses in the whole ring \cite{Scandale2016129}.

\begin{figure}
\includegraphics[width=1\columnwidth]{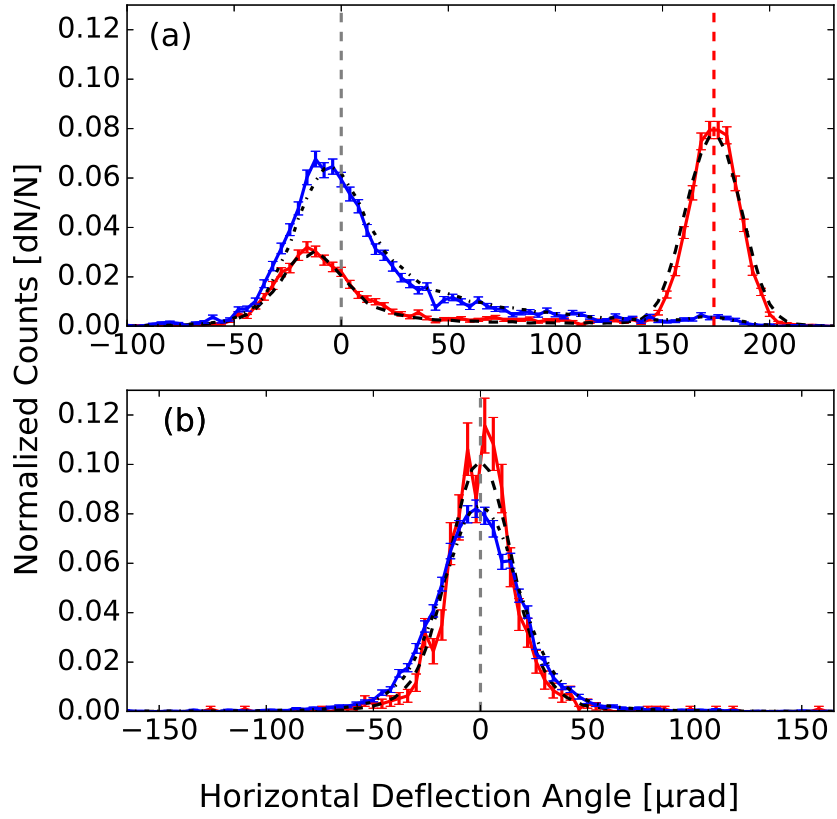}
\caption{\label{fig:geant4} Experimental measurements (red for e$^{+}$ and blue for e$^{-}$) and Geant4 simulations (black dashed line for $e^{+}$ and black dash-dotted line for $e^{-}$) of the deflection-angle distribution in the bent (a) and free (b) directions for  $e^{+}$ and $e^{-}$ beams interacting with the crystal. Only particles with an incoming angle less than half of the critical channeling angle, $\theta_{c}$ ($18.8$ $\mu$rad for $120$ GeV/c $e^{+}$), with respect to the channeling plane have been analyzed.}
\end{figure}

Particles under channeling undergo coherent interactions with the crystalline nuclei of planes or axes. Coherent interactions have been interpreted in terms of a continuous potential by Lindhard \cite{Dansk.Fys.34.14}, i.e. time-reversible particle dynamic is governed by the conservative time-independent electric potential generated by the ordered atomic lattice. However, other than the interaction with a crystal as a whole, particles naturally suffer interactions with individual nuclei and electrons, which may abruptly vary the particle kinetic energy. This latter is called \textit{dechanneling} and is the main limiting factor for applications of channeling.

Experiments with thin crystals allowed to measure the dechanneling intensity for some positive and negative particles. In particular, the comparison of the results obtained for 150 GeV/c negative pions with the experimental data for the dechanneling on atomic nuclei of 400 GeV/c positive particles hinted that the intensity of the two phenomena may have the same magnitude.

\begin{figure}
\includegraphics[width=1\columnwidth]{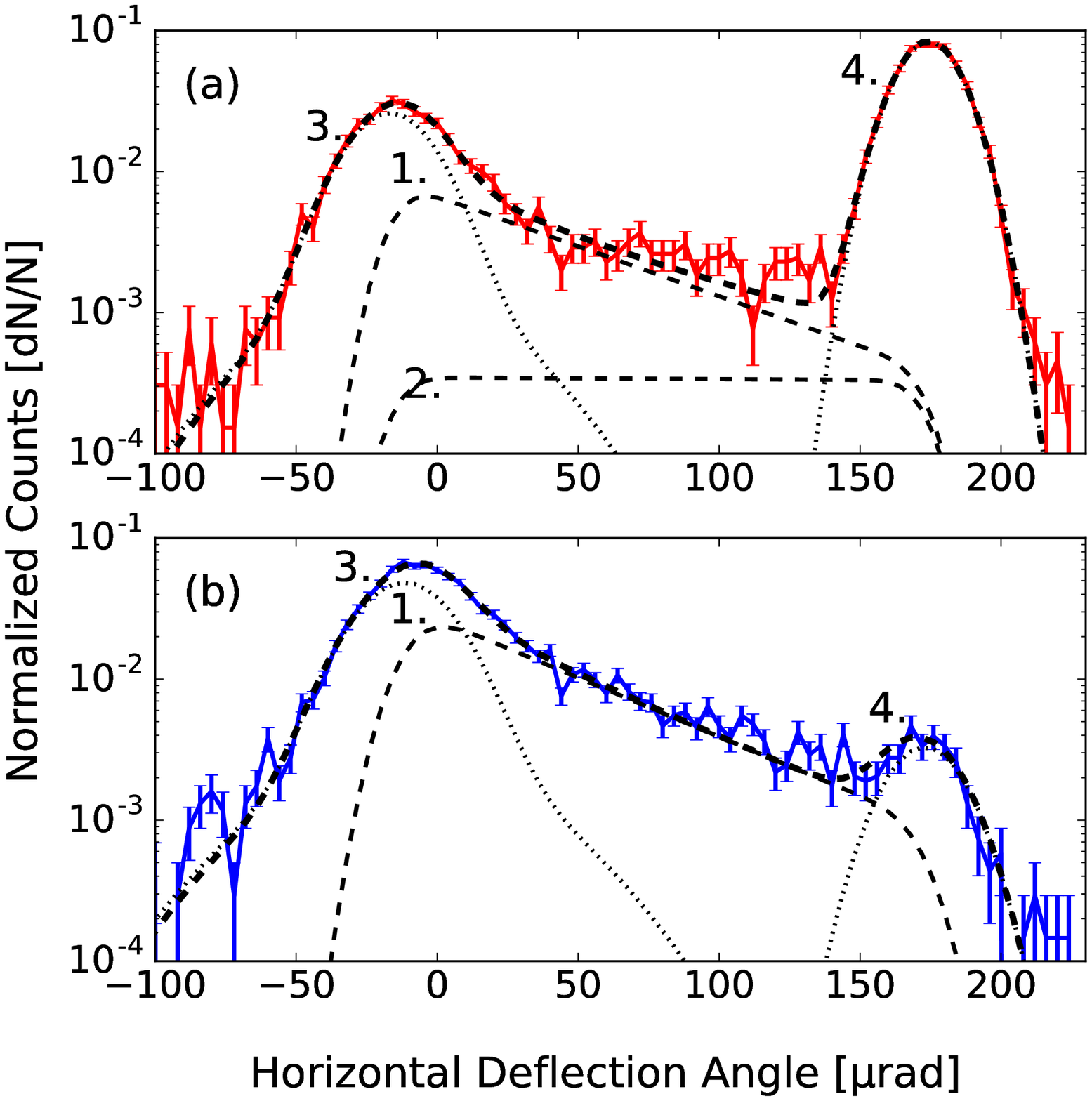}
\caption{\label{fig:comparison} Experimental measurements ((a) for e$^{+}$ and (b) e$^{-}$) and fitted distributions (black dash-dotted for the fitted distribution, dashed for the nuclear (1.) and electronic (2.) dechanneling components and dotted lines for the undeflected (3.) and channeling (4.)) of the deflection-angle distribution in the bent direction for e$^{+}$ and  e$^{-}$ beams interacting with the crystal. Only particles with an incoming angle less than half of the critical channeling angle, $\theta_{c}$ ($18.8$ $\mu$rad for $120$ GeV/c $e^{+}$), with respect to the channeling plane have been analyzed.}
\end{figure}

In this paper we experimentally investigate the nature of the dechanneling process by the same bent Si crystal with a particle and its anti-particle ($e^{-}$ and $e^{+}$) at the same 120 GeV/c beam energy. The experiments were carried out at the H4-SPS line at CERN. 

The motion of channeled particles is affected by incoherent scattering processes with electrons and nuclei that cause the non-conservation of the transverse energy. As a consequence, the transverse energy may exceed the potential well barrier causing the particle to leave the channeling state, i.e., they are dechanneled. Dependently on their transverse energy, particles can either enter or not the nuclear corridor, i.e. the volume of the crystal within which nuclei perform their thermal vibration. The fraction of particles that have sufficient transverse energy to enter the nuclear corridor is $f_{n}$ at the crystal entry face ($z=0$). The remaining fraction is $f_{e}=1-f_{n}$.

Due to strong interaction with nuclei, $f_{n}$ rapidly shrinks, e-folding at distance $l_{n}$ from the surface, which is called the nuclear dechanneling length. In order to estimate the fraction $f_{n}$, the atomic density can be approximated as a Gaussian distribution with standard deviation equal to the atomic thermal vibration amplitude ($u_{t}$) which is $0.075$ ${\AA}$ for Si at $273$ K \cite{Biryukov}. By assuming that the atomic density region where intense multiple scattering occurs extends over $2.5$ $u_{t}$ \cite{Scandale2009129} - the so-called nuclear corridor - and bearing in mind that  (110) interplanar distance is $d_{p}=1.92$ ${\AA}$, $\sim19.5$ $\%$ of particles of a perfectly parallel beam are subject to nuclear dechanneling.

The remaining $f_{e}$ fraction of particles does not initially interact with nuclei and thereby undergoes interaction with electrons only. The interaction strengths with electrons and nuclei are quite different, e.g., for a collimated $400$ GeV/c proton beam interacting with Si (110) crystal, $l_{e}^{(+)}\sim220$ mm \cite{Carrigan}, while $l_{n}^{(+)}\sim1.5$ mm \cite{Scandale2009129}, with  $l_{e}^{(+)}$ being the electronic dechanneling length. The electronic dechanneling length scales proportionally to the particle momentum \cite{Arduini1998325} and was extensively measured for protons \cite{Bak19841,Elishev1979387,Gibson198454}.

In the literature \cite{Scandale2009129,Scandale201370,doi.org/10.1140/epjc/s10052-014-2740-7}, the channeled-particle fraction $f_{ch}^{(+)}$ at depth $z$ in the crystal holds 

\begin{equation}
f_{ch}^{(+)}(z){\approx}f_{n}e^{-z/l_{n}^{(+)}}+f_{e}e^{-z/l_{e}^{(+)}}
\label{eq1}
\end{equation}

where $f_{ch}$ is the fraction of channeled particles and $l_{e}^{(+)}$ is the electronic dechanneling length, i.e. the distances after which a $1/e$ fraction of the initial particles are still under channeling. 

For negatively charged particles, since the minimum of the potential well is located on the atomic planes, the mechanism of electronic dechanneling has a negligible contribution ($f_{e}\sim0$, $f_{n}\sim1$) because all the particles do interact with nuclei \cite{Scandale201370,PhysRevLett.112.135503,PhysRevLett.114.074801}. Therefore, the channeled-particle fraction $f_{ch,-}$ holds

\begin{equation}
f_{ch}^{(-)}(z){\approx}e^{-z/l_{n}^{(-)}}
\label{eq2}
\end{equation}

\begin{figure}
\includegraphics[width=1\columnwidth]{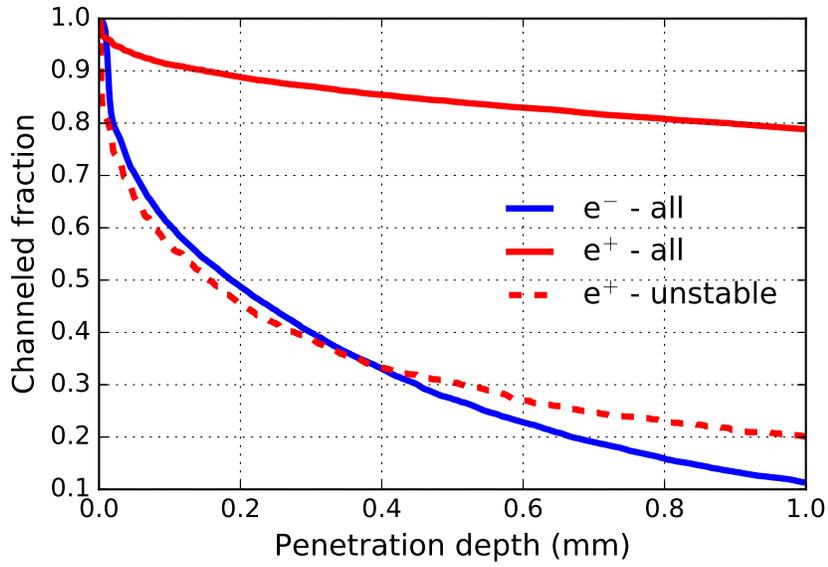}
\caption{\label{fig:strComp} Simulation of the fraction of particles under channeling interacting with a (110) Si straight crystal for 120 GeV/c e$^{-}$ (e$^{-}$ - all), e$^{+}$ (e$^{+}$ - all) and for the fraction of channeled e$^{+}$ that impinge on the crystal close to the atomic planes (e$^{+}$ - unstable).}
\end{figure}

A bent crystal is capable of separating channeled, never-channeled and dechanneled fractions \cite{Scandale2009129,Scandale201370}. Indeed, the channeled fraction is deflected to the nominal crystal bending angle, the never-channeled fraction is only scattered, while the dechanneled fraction results in a deflection angle lower than the nominal crystal bending angle. Therefore, for the measurement of $l_{n}^{(+)}$ and $l_{n}^{(-)}$, a slightly bent thin crystal is the optimal choice, since the three particle fractions can be easily discriminated.  Channeling efficiency decreases as the crystal curvature $1/R$ increases, and vanishes for $R<R_{c}$, $R_{c}$ being the critical radius for channeling \cite{Tsyganov682,Tsyganov684}. The usage of a crystal with $R{\gg}R_{c}$ does not significantly lower the channeling efficiency. In fact, particles dechanneled at a crystal depth $z$ are deflected by an angle $\theta_{z}{\approx}z/R$, thereby a measure of the rate of dechanneled particles as a function of the crystal depth can be inferred \cite{Gibson198454,Scandale2009129}. Moreover, a crystal with $l{\ll}l_{e}$ allows distinguish the nuclear dechanneling length $l_{n}$ for positive particles, since the contribution of the second term in Eq.\ref{eq1} is very small.

The quantities $l_{e}^{(+)}$ and $R_{c}$ scale as particle momentum-velocity $p\beta$ \cite{Biryukov}. For $120$ GeV/c particle momentum, $l_{e}^{(+)}$ is $\approx45$ mm and $R_{c}$ is $\approx0.21$ m. A Si strip that fulfills the requirements $l{\ll}l_{e}$ and $R{\gg}R_{c}$  was manufactured starting from prime-quality wafers. The strip was shaped as a parallelepiped of size 1.99 x 55.0 x 2.01 mm$^{3}$, with $l_{cry}=2.01$ mm being the length along the beam direction, and was bent using a custom-made mechanical device \cite{AfoninU70JETP}, resulting in an anticlastic bending of the (110) planes with $R=(11.5\pm0.5)$ m. The strip bending radius \cite{ipac10tors} was determined by means of interferometric and diffractometric measurements  through a VEECO NT-1100 white-light interferometer and a Panalytical X-Pert MRD-PRO diffractometer, respectively.

The crystal was exposed to a 120 GeV/c e$^{-}$ and e$^{+}$ beams at the H4 CERN-SPS extracted line with $(66\pm2)$x$(97\pm5)$ $\mu$rad$^{2}$ divergence rms. The holder with the crystal was mounted on a two-axis goniometer with an angular resolution of $\sim1$ $\mu$rad. The particle incoming and outgoing angles from the crystal were detected by means of a tracking detectors system \cite{Celano199649}. The standard deviation of $14.6$ $\mu$rad for the angular resolution of the system was verified through Geant4 Monte Carlo simulations \cite{Agostinelli2003250,Allison2016186}. An electromagnetic calorimeter was positioned after the telescope system, allowing the selection of e$^{-}$ and e$^{+}$ and the rejection of muons and hadrons. The strip largest face orthogonal to the $\langle110\rangle$ axis was oriented parallel to the beam direction.

\begin{table}[ht]
\centering
\caption{Fit parameters for the distribution of the deflection angles in the bent direction for $120$ GeV/c $e^{+}$ and $e^{-}$ after the interaction with the $2.01$ mm Si (110) crystal, where $f_{c}$, $\theta_{c}$ and $\sigma_{c}$ and $f_{u}$, $\theta_{u}$ and $\sigma_{u}$ are the efficiency, mean deflection and standard deviation of the channeling and undeflected peaks, $l_{n}$ the nuclear dechanneling length and $f_{n}$ the fraction of particles under nuclear dechanneling. The parameters $A$ and $r$ for the double Gaussian distribution of the undeflected peak were set equal to $0.89$ and $2.3$, as for the misaligned orientation (see \ref{tab:tabY}), and $l_{e}$ was set to $44.5$ mm \cite{Biryukov}.}
\label{tab:tabX}
  \begin{tabular}{ c | c c c c c c c c c }
    Type   &  Particle & $f_{u}$  & $\theta_{u}$   & $\sigma_{u}$   & $f_{c}$   & $\theta_{c}$  & $\sigma_{c}$  & $l_{n}$ & $f_{n}$ \\
       &   & $\%$  & $\mu$rad  & $\mu$rad   & $\%$  & $\mu$rad  & $\mu$rad  & mm & $\%$  \\
     \hline
   Data        & $e^{+}$ & $23\pm1$ & $-17\pm1$ & $8.2\pm0.1$        & $54\pm2$ & $174\pm2$ & $6.3\pm0.1$ & $0.7\pm0.3$ & $30\pm10$\\
    Geant4   & $e^{+}$ & $25\pm1$ & $-15\pm1$ & $8.6\pm0.1$       & $59\pm2$ & $174\pm2$ & $7.2\pm0.1$ & $0.7\pm0.2$ & $21\pm5$\\
\hline\hline    
    Data       & $e^{-}$ & $41\pm2$ & $-11\pm1$ & $9.1\pm0.2$       & $2\pm2$ & $173\pm2$ & $7.5\pm0.7$ & $0.6\pm0.1$ & $100$ \\
    Geant4   & $e^{-}$& $33\pm2$ & $-10\pm1$ & $9.1\pm0.1$   & $2\pm2$ & $173\pm2$ & $8.5\pm0.3$ & $0.6\pm0.2$ & $100$ \\
 \end{tabular}
\end{table}

Figure \ref{fig:geant4}.a shows the experimental distributions of the deflection-angle under channeling for $e^{-}$ and $e^{+}$. Since particles undergo channeling when the angle $\theta$ between their direction and the crystal planes is smaller than the critical angle for channeling, $\theta_{c}$ ($18.8$ $\mu$rad for $120$ GeV/c $e^{+}$), only the particles with $\theta<\theta_{c}/2$ were analyzed.

The analysis of the distributions was carried out using the fitting procedure described in Ref. \cite{PhysRevAccelBeams.19.071001} for $e^{-}$. The probability distribution of the dechanneling particles ($dP_d/d{\theta}(\theta)$) is

\begin{equation}
\frac{df_d}{d\theta}(\theta,\theta_d) = \frac{1-f_{u}}{2\theta_{d}}e^{\frac{\sigma_{u}^2}{2\theta_d^2}+\frac{\theta_{c}}{\theta_d}-\frac{\theta}{\theta_d}}
\left( \erf \left( \frac{\theta_{u}-\Delta\theta}{\sqrt(2)\sigma_{u}}\right) - \erf \left( \frac{\theta_{c}-\Delta\theta}{\sqrt(2)\sigma_{u}}\right) \right)
\label{eq:dechNeg}
\end{equation}

where $f_u$ is the fraction of particles in the undeflected peak, $\sigma_u$ and $\theta_u$ the standard deviation and the mean of the distribution for the undeflected peak, $\theta_c$ the mean of the distribution for the channeling peak, $\theta_d=l_n/l_{cry}\theta_c$ and $\Delta\theta=\theta-\sigma_u^2/\theta_d$. Figure \ref{fig:comparison}.b shows the fitted distribution over the experimental one and the fit parameters are summarized in Tab. \ref{tab:tabX}. The channeling efficiency is $(2\pm2)$ $\%$ and the dechanneling length is $(0.6\pm0.1)$ mm.

The measured dechanneling length is shorter than the crystal length. As a consequence, the fraction of particles dechanneled due to electronic dechanneling may be visible. Therefore, the probability distribution of the dechanneling tail for $e^{+}$ becomes the sum of two terms:

\begin{equation}
f_{e}\frac{df_{d}}{d\theta}(\theta,\theta_{e}) + f_{n}\frac{df_{d}}{d\theta}(\theta,\theta_{n})
\end{equation}

where $\theta_{e}=l_e/l_{cry}\theta_c$ and $\theta_{n}=l_n/l_{cry}\theta_c$, $f_{n}$ and $f_{e}$ being the fraction of channeled particles under nuclear and electronic dechanneling, respectively. Figure \ref{fig:comparison}.a shows the fitted distribution over the experimental one and the fit parameters are summarized in Tab. \ref{tab:tabX}. The channeling efficiency is $(54\pm2)$ $\%$, $l_n^{(+)}=(0.7\pm0.3)$ mm and $f_{n}^{(+)}=(30\pm10)$ $\%$. The same fit procedure was repeated for the case of electrons, resulting in a $f_{n}^{(-)}=(0.0\pm0.1) \%$, which is consistent with the initial supposition that negative particles are subject to nuclear dechanneling only. As previously noted, the rate of nuclear dechanneling depends on incoherent interactions with atomic nuclei, which are similar for positive and negative particles. In fact, measurements showed that $l_{n}$ does not significantly vary with particle charge at all.

\begin{figure}
\includegraphics[width=1\columnwidth]{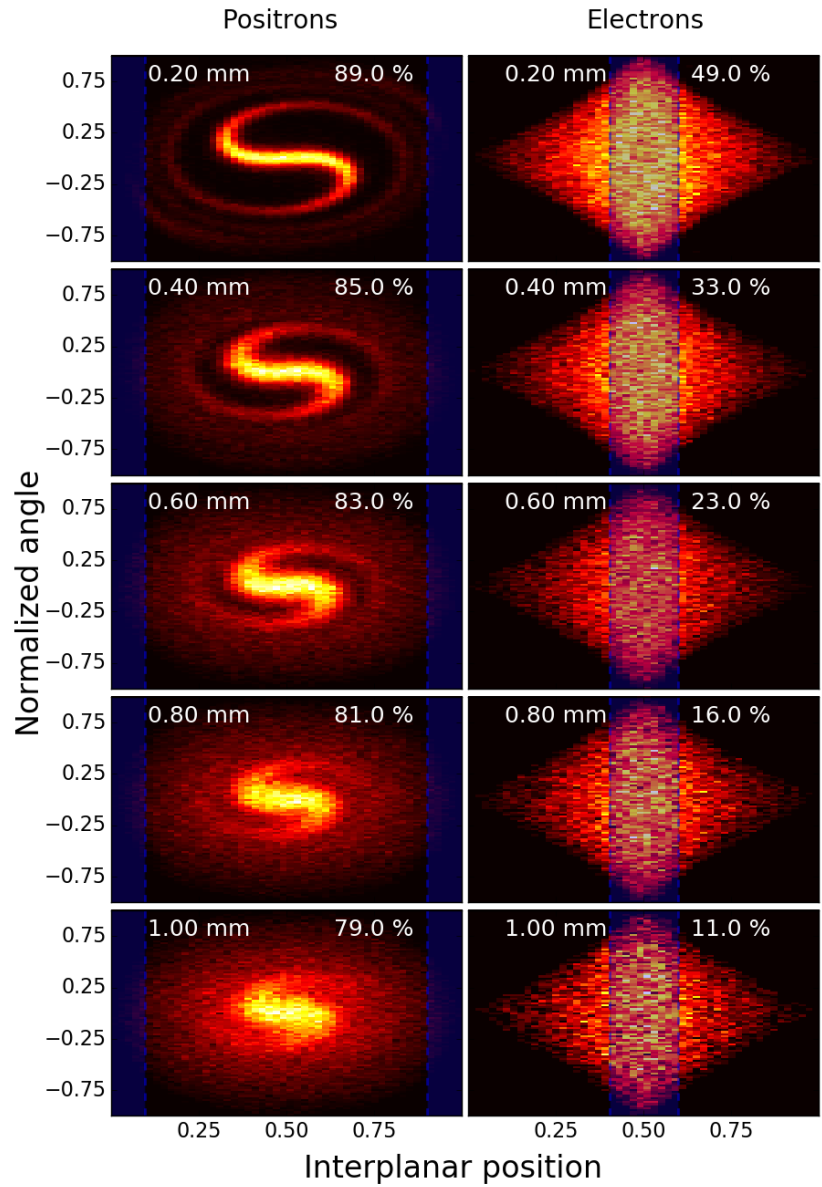}
\caption{\label{fig:phaseSpace} Evolution of the phase space for a collimated beam of e$^{+}$ and e$^{-}$ interacting with a Si (110) straight crystal. The full sequence is available as a supplementary material. Top right and top left of each figure show the penetration depth at which the snapshots was recorded and the fraction of particles under channeling, i.e. the channeling efficiency.}
\end{figure}

Figure \ref{fig:geant4}.b shows the experimental distributions of the deflection-angle under channeling for $e^{-}$ and $e^{+}$ for the free direction, i.e. the direction not bent. The analysis of the distributions was carried out using the fitting procedure described in Ref. \cite{PhysRevAccelBeams.19.071001} and the fit parameters are summarized in Tab. \ref{tab:tabY}. The scattering in the free plane for negative particles is stronger for the particles under channeling than for those not aligned with the crystal planes, while for positive particles the opposite occurs. Indeed for positive particles the largest fraction of particles does not interact with the nuclei, reducing the probability of incoherent scattering.

\begin{table}[ht]
\centering
\caption{Fit parameters for the distribution of the deflection angles in the free direction after the interaction of $120$ GeV/c $e^{+}$ and $e^{-}$ with the $2.01$ mm Si (110) crystal under channeling. The distribution is the sum of two Gaussians, where $A$ is the constant factor of the first Gaussian ($1-A$ for the second Gaussian) and $r$ is the ratio between the standard deviations of the second Gaussian and the one of the first Gaussian. The same fit was carried out for particles not aligned with the crystal planes (Not aligned) for both $e^{-}$ and $e^{-}$.}
\label{tab:tabY}
  \begin{tabular}{ c | c | c c c c }
    Type   & Condition & Particle & $\sigma$       & $A$                       & $r$ \\
       &   & $\mu$rad        &                        &  \\
     \hline
   Data & Channeling & $e^{+}$   & $7.6\pm0.4$   & $0.88\pm0.03$ & $3.3\pm0.7$\\
    Geant4   & Channeling & $e^{+}$   & $8.6\pm0.1$   & $0.91\pm0.02$ & $2.7\pm0.1$\\
\hline\hline    
    Data & Channeling & $e^{-}$    & $10.3\pm0.2$ & $0.87\pm0.02$ & $2.5\pm0.1$\\
    Geant4   & Channeling & $e^{-}$    & $10.2\pm0.1$ & $0.89\pm0.01$ & $2.6\pm0.1$\\
\hline\hline    
Data & Not Aligned & $e^{+}/e^{-}$  & $8.8\pm0.1$ & $0.89\pm0.02$ & $2.3\pm0.1$\\
 \end{tabular}
\end{table}

Monte Carlo simulations were carried out using the Geant4 toolkit \cite{Agostinelli2003250,Allison2016186}. The experimental setup at the H4-SPS area is reproduced in the simulation in order to take into account the error due to the finite resolution of the telescope. Channeling is implemented via an updated version of the Geant4 channeling package \cite{s10052-014-2996-y}. The package does not take into account coherent radiation processes. The results are shown in Fig. \ref{fig:geant4} and the fit parameters are summarized in Tabs. \ref{tab:tabX} and Tab. \ref{tab:tabY}. Simulations show a good agreement with experimental data for both positive and negative particles.

The availability of a Monte Carlo code for the simulations of the coherent phenomena allows to having an insight into the dechanneling mechanism. Indeed, other than comparing the deflection distribution at the exit of the crystal, the evolution of the beam phase space into the crystal can be studied. Figure \ref{fig:strComp} shows the evolution of the fraction of channeled particles as a function of the penetration depth into a straight crystal of 1 mm. In the simulation 120 GeV/c e$^{-}$ and e$^{+}$ collimated beams impinge on a Si (110) crystal. The simulations were worked out via the DYNECHARM++ code \cite{Bagli2013124,PhysRevE.81.026708}. As can be noticed, the fraction of particles in unstable channeling condition that impinge on the crystal close to the atomic planes, i.e. at a distance less than $2.5$ times the amplitude of atomic thermal vibration, decreases as the fraction of channeled particles for a e$^{-}$ beam. On the other hand, the channeling efficiency of the whole e$^{+}$ beam remains higher than $80$ $\%$.

Figure \ref{fig:phaseSpace} shows five snapshots of the evolution of the phase space for a perfectly collimated beam of e$^{+}$ and e$^{-}$ interacting with a Si (110) straight crystal for a lively representation of the particle dynamics (evolution sequence is available as supplementary material). The evolution of the e$^{+}$ particles under stable channeling condition, i.e. that oscillate far from atomic plane, maintains a coherent pattern in the phase space for a period much longer than the e$^{-}$ particles. For positive particles the length for which the confined channeled particles are randomly distributed in the phase-space spot is similar to the $l_{n}^{(+)}$, while the distribution of negative particles is immediately randomized approximately after a single oscillation period due to the strong interaction with atomic nuclei.

In summary, the nuclear dechanneling lengths of 120 GeV/c $e^{-}$ and $e^{+}$ were measured. State-of-the-art slightly bent Si crystal were adopted to separate channeled, unchanneled and dechanneled fractions, resulting in the capability of measuring the rate of incoherent interactions with nuclei. We found out that the channeling efficiency is different, $2\pm2$ $\%$ for $e^{-}$ and $54\pm2$ $\%$ for $e^{+}$, while the nuclear dechanneling length is comparable, $l_{n}^{(-)}$ and $l_{n}^{(+)}$ being $(0.6\pm0.1)$ mm and $(0.7\pm0.3)$, i.e. the experimental proof that the nuclear dechanneling length does not depend on particle charge. Such result is fundamental for the design and fabrication of crystals suitable for the manipulation of both positive and negative particles. Moreover, the already existing experimental studies for negative particles may be extended to positive particles and vice versa. As an example, the nuclear dechanneling length for positrons at MAMI or SLAC energies may be inferred from the measured dechanneling length of electrons at the same energies, or the nuclear dechanneling length of anti-proton at 400 GeV/c from the measured dechanneling length of protons.

We acknowledge partial support by the Instituto Nazionale di Fisica Nucleare under the CHANEL and GECO projects and by the H2020 project AIDA-2020, GA no. 654168. Authors acknowledge the help and support of SPS-CERN coordinator and staff.

\bibliography{biblio}
\bibliographystyle{spphys}

\end{document}